\documentstyle[prd,aps]{revtex}

\begin{document}
\draft

\twocolumn[
\hsize\textwidth\columnwidth\hsize\csname
@twocolumnfalse\endcsname

\title{\noindent {\small USC-97/HEP-B3\hfill \hfill hep-th/9705205}
\newline
{\small LPTENS-97/25\hfill CERN-TH/97-108}\\
String and Particle with Two Times}
\author{Itzhak Bars and Costas Kounnas}
\address{\it Theory Division, CERN, CH-1211 Geneva 23, Switzerland}
\date{May 27, 1997}
\maketitle
\begin{abstract}
An action for a string and a particle with two timelike dimensions
is proposed and analyzed. Due to new gauge symmetries and associated
constraints, the motion of each system in the background of the other is
equivalent to effective motion with a single timelike dimension. The quantum
constraints are consistent only in critical dimensions. For the bosonic
system in flat spacetime the critical dimension is 27 or 28, with signature
(25,2) or (26,2), depending on whether the particle is massive or massless
respectively. For the supersymmetric case the critical dimensions are 11 or
12, with signature (9,2) or (10,2), under the same circumstances. Generalizations to multi particles, strings and p-branes are outlined.  
\end{abstract}
\pacs{PACS: 11.17.+y, 02.40.+m, 04.20.Jb } 
\vskip2pc
]

\section{Classical particles and strings}

The idea that the fundamental theory may be formulated in twelve or more
dimensions has been receiving increased attention \cite{duff} - \cite{towns}
. It has become apparent that some of the extra dimensions are timelike, and
thus the issues associated with more than one timelike dimensions must be
addressed seriously. As a first step toward theories with two or more
timelike dimensions, in which the traditional problems are overcome, we have
proposed a set of gauge symmetries and associated constraints, as well as a
cosmological scenario \cite{twotimes}. As an example we formulated an action
principle for two particles which move freely except for a global constraint
on each other's momenta. In this paper we generalize this type of action
principle by discussing the example of a string and a particle in detail,
and then showing how to apply the same methods to more particles, strings
and p-branes. We discuss the quantum constraints, the emergence of critical
dimensions, and the quantum consistent sectors. The supersymmetric
generalization is outlined.

\subsection{Reformulation of the two particles}

Consider two particles described by their worldlines $x_1^\mu \left( \tau
\right) ,x_2^\mu \left( \tau \right) $. In our previous work \cite{twotimes}
we presented an action with appropriate gauge invariances that produced the
following constraints for the momenta of the two particles 
\begin{equation}
p_1^2+m_1^2=0,\quad p_2^2+m_2^2=0,\quad p_1\cdot p_2=0.  \label{system}
\end{equation}
The two particles move freely, except for the mutual constraint $p_1\cdot
p_2=0.$ Two orthogonal timelike momenta cannot exist in a space with a
single timelike dimension. The extra constraint was the key for two timelike
dimensions and their interpretation given in \cite{twotimes}. In this
section we would like to give another formulation of the action that leads
to the same results. The new formulation is better adapted to
generalizations to strings and p-branes.

We consider the following action for two particles

\begin{eqnarray}
S &=&S_1(x_1,A_1,e_1,\lambda _2)+S_2(x_2,A_2,e_2,\lambda _1)+\lambda _1^\mu
\,\lambda _2^\nu \,\eta _{\mu \nu }  \nonumber \\
S_1 &=&\frac 12\int_0^Td\tau \,\left[ e_1^{-1}\left( \partial _\tau x_1^\mu
-\lambda _2^\mu A_1\right) ^2-e_1m_1^2\right]  \label{partaction}
\end{eqnarray}
and similarly for $S_2$, where $x_i^\mu \left( \tau \right) ,A_i\left( \tau
\right) ,e_i\left( \tau \right) $ are functions of $\tau $ while $\lambda
_i^\mu $ are independent of $\tau .$ Note that $\lambda _2^\mu $ appears in $
S_1$ and $\lambda _1^\mu $ appears in $S_2$. As we will see, $\lambda _2^\mu 
$ is determined in terms of canonical variables that belong to particle \#2,
so that the presence of particle \#2 influences particle \#1, and vice
versa. In the path integral we integrate over the $\lambda _i^\mu $ as well
as the other fields. Therefore, in the classical theory we minimize the
action with respect to the $\lambda _i^\mu $ as well as the other fields.

The $\tau $ reparametrization invariance of $S_i(x_i,A_i,e_i,\lambda
_{i^{\prime }})$ is independent for each $i$ (we denote $1^{\prime }=2$ and $
2^{\prime }=1$), hence there are two reparametrizations that eventually
allow the two gauge choices $e_i\left( \tau \right) =1.$ The equations of
motion for the $e_i$ lead to two constraints 
\begin{equation}
p_i^2+m_i^2=0,
\end{equation}
where 
\begin{equation}
p_i^\mu =e_i^{-1}\left( \partial _\tau x_i^\mu -\lambda _{i^{\prime }}^\mu
A_i\right) ,\quad \partial _\tau p_i^\mu =0
\end{equation}
are the canonical momenta, which are conserved according to the equations of
motion for $x_i^\mu .$

The action is gauge invariant under the following two gauge transformations
with parameters $\Lambda _i\left( \tau \right) $ 
\begin{equation}
\delta x_i^\mu \left( \tau \right) =\lambda _{i^{\prime }}^\mu \Lambda
_i\left( \tau \right) ,\quad \delta A_i\left( \tau \right) =\partial _\tau
\Lambda _i\left( \tau \right) .
\end{equation}
The covariant derivatives $\partial _\tau x_i^\mu -\lambda _{i^{\prime
}}^\mu A_i$ may be seen as arising from the gauging of an Abelian subgroup
in the spirit of gauged Wess-Zumino-Witten (WZW) models. Because of these
gauge invariances, the equations of motion for $A_i$ lead to two other
constraints that help remove degrees of freedom 
\begin{equation}
\lambda _{i^{\prime }}\cdot p_i=0.  \label{constr}
\end{equation}

Finally we come to the role of the coupling term $\lambda _1\cdot \lambda _2$
in the action. If it were not for this coupling term there would be two
independent Lorentz symmetries, one for each particle. However, because of
this coupling there is a single Lorentz symmetry, in a $d$-dimensional space
with a metric $\eta _{\mu \nu }$ whose signature will be determined by the
solution of all the constraints. The equations of motion for the $\lambda _i$
give 
\begin{equation}
\lambda _1^\mu =\int_0^Td\tau \,A_1p_1^\mu ,\quad \lambda _2^\mu
=\int_0^Td\tau \,A_2p_2^\mu \,\,.
\end{equation}
Since the momenta are conserved one finds 
\begin{equation}
\lambda _1^\mu \sim p_1^\mu ,\quad \lambda _2^\mu \sim p_2^\mu .
\end{equation}
Therefore, the two constraints in (\ref{constr}) reduce to the single
constraint 
\begin{equation}
p_1\cdot p_2=0.
\end{equation}

We have demonstrated that the new action in (\ref{partaction}) reproduces
the same system of constraints in (\ref{system}) given by our old action.
Hence, we need to introduce two time-like dimensions and interpret them as
in our previous work. This reformulation is more elegant and permits
generalizations to strings and p-branes, as discussed below.

\medskip

\subsection{A string and a particle}

Consider a string and a particle described by the a worldsheet $x_1^\mu
\left( \tau ,\sigma \right) $ and a worldline $x_2^\mu \left( \tau \right) $
respectively, and introduce the action 
\begin{eqnarray}
S &=&S_1(x_1,A_{1m},g_{mn},\lambda _2)+S_2(x_2,A_2,e_2,\lambda _1)+\lambda
_1\cdot \lambda _2\,  \nonumber \\
S_1 &=&\frac 12\int_0^Td\tau \int d\sigma \,\,\sqrt{-g}g^{mn}\left( \partial
_mx_1^\mu -\lambda _2^\mu A_{1m}\right)   \nonumber \\
&&\,\,\,\,\,\,\,\,\,\,\,\,\,\,\,\,\,\,\,\,\,\,\,\,\,\,\,\,\,\,\,\,\,\,\,\,\,
\,\,\,\,\,\,\,\,\,\,\times \left( \partial _nx_1^\nu -\lambda _2^\nu
A_{1n}\right) \eta _{\mu \nu }  \label{partstr} \\
S_2 &=&\frac 12\int_0^Td\tau \,\left[ e_2^{-1}\left( \partial _\tau x_2^\mu
-\lambda _1^\mu A_2\right) ^2-e_2m_2^2\right]   \nonumber
\end{eqnarray}
The two actions $S_{1,2}$ are invariant under independent
reparametrizations, hence one can choose the usual conformal gauge for the
string $\sqrt{-g}g^{mn}=\eta ^{mn},$ and $e_2=1$ for the particle, and
obtain the following constraints from the equations of motion of$
\,\,\,g^{mn},e_2$ respectively 
\begin{equation}
\left( D_{\pm }x_1^\mu \right) ^2=0,\quad p_2^2+m_2^2=0,  \label{con}
\end{equation}
where 
\begin{eqnarray}
D_{\pm }x_1^\mu  &=&\left( \partial _{\pm }x_1^\mu -\lambda _2^\mu A_{1\pm
}\right) ,\quad  \\
p_2^\mu  &=&e_2^{-1}\left( \partial _\tau x_2^\mu -\lambda _1^\mu A_2\right)
,  \nonumber
\end{eqnarray}
and the $\partial _{\pm }$ derivatives are with respect to the lightcone
variables $\sigma ^{\pm }\equiv \tau \pm \sigma .$ The equations of motion
for $x_1^\mu \left( \tau ,\sigma \right) ,x_2^\mu \left( \tau \right) ,$ in
the gauges we have chosen, are 
\begin{equation}
\partial _{+}\left( D_{-}x_1^\mu \right) +\partial _{-}\left( D_{+}x_1^\mu
\right) =0,\quad \partial _\tau p_2^\mu =0.
\end{equation}
As in the two particle case, there are additional gauge invariances, which
may be understood in the spirit of gauged WZW models 
\begin{eqnarray}
\delta _1x_1^\mu  &=&\lambda _2^\mu \Lambda _1\left( \tau ,\sigma \right)
,\quad \delta _1A_{1m}=\partial _m\Lambda _1\left( \tau ,\sigma \right) ,
\label{gaugesym1} \\
\delta _2x_2^\mu  &=&\lambda _1^\mu \Lambda _2\left( \tau \right) ,\quad
\,\quad \delta _2A_2=\partial _\tau \Lambda _2\left( \tau \right) , 
\nonumber
\end{eqnarray}
and these explain the structure of the covariant derivatives ($
g_{mn},e_2,\lambda _{1,2}^\mu $ are invariant under $\delta _{1,2})$. The
equations of motion for $A_{1\pm },A_2$ give the constraints associated with
these gauge invariances 
\begin{equation}
\lambda _2\cdot D_{\pm }x_1=0,\quad \lambda _1\cdot p_2=0.  \label{constr2}
\end{equation}
Finally, the equations of motion for $\lambda _{1,2}^\mu $ give 
\begin{eqnarray}
\lambda _1^\mu  &=&\int_0^Td\tau \,\int d\sigma \,\,\left( D_{+}x_1^\mu
A_{1-}+D_{-}x_1^\mu A_{1+}\right) , \\
\lambda _2^\mu  &=&\int_0^Td\tau \,A_2p_2^\mu \,,\quad   \nonumber
\end{eqnarray}
Using the constraints (\ref{constr2}) we deduce 
\begin{equation}
\lambda _1\cdot \lambda _2=0.
\end{equation}
Since $p_2^\mu $ is conserved, one finds $\lambda _2^\mu \sim p_2^\mu .$

From the first equation in (\ref{constr2}) we deduce $A_{1\pm }=\frac 1{
\lambda _2^2}\lambda _2\cdot \partial _{\pm }x_1$ which allows us to write 
\begin{equation}
D_{\pm }x_1^\mu =\partial _{\pm }x_1^\mu -\frac 1{p_2^2}p_2^\mu \,\,p_2\cdot
\partial _{\pm }x_1\,,
\end{equation}
which solves the constraint $\lambda _2\cdot D_{\pm }x_1=p_2\cdot D_{\pm
}x_1=0$ explicitly, provided $p_2^2\neq 0$. From this form one sees that the
component of $x_1^\mu $ that is parallel to $p_2^\mu $ drops out of the
string system when $p_2^2\neq 0$. Using the $\Lambda _1\left( \tau ,\sigma
\right) $ gauge invariance one may choose the gauge 
\begin{equation}
p_2\cdot x_1\left( \tau ,\sigma \right) =2c_1p_2^2\tau =c_1p_2^2\left(
\sigma ^{+}+\sigma ^{-}\right) ,  \label{gauge}
\end{equation}
where $c_1$ is a constant. In this gauge one has 
\begin{equation}
D_{\pm }x_1^\mu =\partial _{\pm }x_1^\mu -c_1p_2^\mu \,,\quad \,A_{1\pm
}\left( \tau ,\sigma \right) =c_1\sqrt{p_2^2/\lambda _2^2},  \label{dpdm}
\end{equation}
which leads to the simplification 
\begin{equation}
\lambda _1^\mu =c_1\sqrt{p_2^2/\lambda _2^2}\int_0^Td\tau \,\int d\sigma
\,\,\left( \partial _\tau x_1^\mu -2c_1p_2^\mu \right) .
\end{equation}
The canonical momentum density for the string is $D_\tau x_1^\mu =\partial
_\tau x_1^\mu -A_{1\tau }\lambda _2^\mu =\partial _\tau x_1^\mu -2c_1p_2^\mu
. $ Inserting this in the equation above, one finds that $\lambda _1^\mu $
is proportional to the {\it total conserved} momentum of the string \# 1 
\begin{equation}
p_1^\mu =\int d\sigma \,\,\left( \partial _\tau x_1^\mu -2c_1p_2^\mu \right) ,
\end{equation}
and that it is orthogonal to the momentum of particle \# 2. Since both $
\lambda _1^\mu $ and $p_1^\mu $ are gauge independent quantities, their
relation which was derived in a specific gauge, is also gauge invariant.
Hence we have deduced that 
\begin{equation}
\lambda _1^\mu \sim \,p_1^\mu ,\quad \lambda _2^\mu \sim p_2^\mu ,\quad
p_1\cdot p_2=0,  \label{lambdas}
\end{equation}
just as in the case of two particles of the previous section.

In the lightlike case $\lambda _2^2=0=p_2^2$ (which is consistent only if $
m_2=0$), the action $S_1\,$has no $A_{1+}A_{1-}$ term, and acquires an
additional gauge symmetry beyond (\ref{gaugesym1}) 
\begin{equation}
\delta _3x_1^\mu \left( \tau ,\sigma \right) =0,\quad \delta _3A_{1\pm }=\pm
\partial _{\pm }\Lambda _3\left( \tau ,\sigma \right) .  \label{three}
\end{equation}
The constraint $p_2\cdot D_{\pm }x_1=p_2\cdot \partial _{\pm }x_1=0$
eliminates a component of $x_1^\mu \left( \tau ,\sigma \right) $ not
parallel to $p_2^\mu $ (this is consistent with the gauge choice (\ref{gauge}
) although in this case it follows from the constraint). The extra $\Lambda
_3\left( \tau ,\sigma \right) $ gauge symmetry can be used to gauge fix $
A_{1\pm }$ to $A_{1\pm }=\partial _{\pm }\gamma \left( \tau ,\sigma \right) $
and then use the gauge symmetry (\ref{gaugesym1}) to fix $\gamma \left( \tau
,\sigma \right) $ so that $D_{\pm }x_1$ and $A_{1\pm }$ take the form in ( 
\ref{dpdm}). In this way all the results above, including (\ref{lambdas}),
apply in the massless particle case as well, and this may be understood as
the limit in which $p_2^2/\lambda _2^2$ remains finite. However, in the
background of the massless particle, two string components, rather than only
one, are eliminated by the gauge invariances: $\partial _{\pm }x_1^\mu
-c_1p_2^\mu $ has no components along the lightlike $p_2^\mu ,$ and $
p_2\cdot \partial _{\pm }x_1=0$, whereas for the massive particle these two
conditions correspond to one and the same component. This kind of phenomenon
happened also in the massless limit of the two particle case, as explained
in \cite{twotimes}.

The equation of motion for the string is easily solved since it has the free
string form 
\begin{equation}
\partial _{+}\partial _{-}x_1^\mu =0.
\end{equation}
As usual, the general solution is given in terms of left and right movers 
\begin{eqnarray*}
x_{1\mu } &=&x_{1\mu }^{\left( +\right) }\left( \sigma ^{+}\right) +x_{1\mu
}^{\left( -\right) }\left( \sigma ^{-}\right) +c_1p_{2\mu }\tau \\
x_{1\mu }^{\left( \pm \right) }\left( \sigma ^{\pm}\right) &=&\frac 12\left(
q_{1\mu }+\frac{\sigma ^{\pm }}{2\pi }p_{1\mu }\right) -i\sum_{n\neq 0}\frac 
1n\alpha _{n\mu }^{\left( \pm \right) }\,\,e^{in\sigma ^{\pm }}
\end{eqnarray*}
The term proportional to $c_1p_2^\mu \tau $ is added to be consistent with
the definition of the total string momentum $p_1^\mu $ that followed from
the canonical formalism. The canonical pair is $(q_{1\mu },p_{1\mu })$,
while the $\alpha _{n\mu }^{\left( \pm \right) }$ have the usual string
oscillator commutation rules.$\,$

The equations of motion for particle \#2 are also solved easily since $
p_2^\mu $ is conserved. The constraint in eq.(\ref{constr2}), together with
the definition of $p_2^\mu $ in terms of the velocity $\partial _\tau
x_2^\mu $ give $A_2\left( \tau \right) =\frac 1{\lambda _1^2}\lambda _1\cdot
\partial _\tau x_2,$ so that 
\begin{equation}
p_2^\mu =\partial _\tau x_2^\mu -p_1^\mu \frac 1{p_1^2}p_1\cdot \partial
_\tau x_2.
\end{equation}
Using the $\Lambda _2\left( \tau \right) $ gauge invariance one can choose
the gauge 
\begin{equation}
p_1\cdot \partial _\tau x_2\left( \tau \right) =c_2p_1^2  \label{gauge2}
\end{equation}
where $c_2$ is a constant. In this gauge the conserved momentum of particle
\#2 and $A_2$ become 
\begin{equation}
p_2^\mu =\partial _\tau x_2^\mu -c_2p_1^\mu ,\quad A_2=c_2\sqrt{
p_1^2/\lambda _1^2}
\end{equation}
This form is valid for the massive as well as massless string states (i.e. $
p_1^2/\lambda _1^2$ finite as $p_1^2\rightarrow 0,\,\,$as above). The
solution of the particle equation is 
\begin{equation}
x_2^\mu \left( \tau \right) =\left( p_2^\mu +c_2p_1^\mu \right) \tau
+q_2^\mu ,
\end{equation}
showing that it moves as a free particle, except for the orthogonality
constraint $p_1\cdot p_2=0.$ The canonical pair is $(q_2^\mu ,p_2^\mu ).$

By reexamining the equations for $\lambda _{1,2}^\mu $ one finds that 
\begin{equation}
\lambda _1^\mu =Ta_1p_1^\mu ,\quad \lambda _2^\mu =Ta_2p_2^\mu ,  \nonumber
\end{equation}
where $a_{1,2}$ are constant zero modes of the gauge fields that have
survived in the general solution 
\begin{equation}
A_{1\pm }\left( \tau ,\sigma \right) \equiv a_1,\quad A_2\left( \tau \right)
=a_2,
\end{equation}
and that the two constants $c_{1,2}$ are equal and given by $
c_1=c_2=Ta_1a_2. $

\section{Quantization and critical dimensions}

In the previous section it was shown that the particle and string systems
move as free systems except for a set of constraints. The canonical degrees
of freedom $(q_2^\mu ,p_2^\mu ),\,(q_1^\mu ,p_1^\mu ),\,\alpha _{n\mu }^{\pm
}$ satisfy the first class constraints that follow from eqs.(\ref{con},\ref
{constr2},\ref{lambdas}) 
\begin{eqnarray}
m_2 &\neq &0:\,\Phi \equiv p_2^2+m_2^2=0,\,\,L_n^{\pm }=0,\,\,\,J_0^{\pm
}\equiv p_2\cdot p_1=0,  \nonumber \\
m_2 &=&0:\,\Phi \equiv p_2^2=0,\,\,L_n^{\pm }=0,\,\,\,J_n^{\pm }\equiv
p_2\cdot \alpha _n^{\pm }=0,  \label{covco}
\end{eqnarray}
where the Virasoro operators are 
\begin{eqnarray}
m_2 &\neq &0:L_n^{\pm }=\frac 12\sum_{m=-\infty }^\infty \alpha
_{n-m}^{\left( \pm \right) \mu }\cdot \alpha _m^{\left( \pm \right) \nu
}\left( \eta _{\mu \nu }-\frac{p_{2\mu }\,p_{2\nu }}{p_2^2}\right)  
\nonumber \\
m_2 &=&0:L_n^{\pm }=\frac 12\sum_{m=-\infty }^\infty \alpha _{n-m}^{\left(
\pm \right) \mu }\cdot \alpha _m^{\left( \pm \right) \nu }\,\,\eta _{\mu \nu
}  \label{covvir}
\end{eqnarray}
$_{}$We have used $\alpha _{0\mu }^{\pm }=p_{1\mu }$, therefore the
constraint $J_0^{\pm }\equiv p_1\cdot p_2=0$ is included above for $m_2=0$
as well. The second term in $L_n^{\pm }$ for $m_2\neq 0$ could be dropped
since $p_2\cdot \alpha _n^{\pm }=0$ was taken as a gauge choice, which may
be added as an additional constraint $J_n^{\pm }\equiv p_2\cdot \alpha
_n^{\pm }=0$ for the massive case as well. This may be done classically as
well as in a lightcone quantization, as we will see below. However, this
gauge choice becomes a second class constraint in the massive case $m_2\neq
0,$ that is harder do deal with in a covariant quantization (since it has an
anomaly proportional to $m_2^2)$ 
\begin{equation}
\left[ J_n^{\pm }\,,\,J_m^{\pm }\,\right] =n\,\Phi \delta
_{n+m}-n\,m_2^2\delta _{n+m}.  \label{anom}
\end{equation}
As we will see, this will play a role in the covariant quantization of the
system and in the determination of the critical dimension.

It is evident from the analogous two particle problem that there is no
solution (either classical or quantum) unless there are two or more timelike
dimensions. Assuming two timelike dimensions the signature of the space is
given by $\eta _{\mu \nu }=diag\left( -1,-1,1,\cdots ,1\right) $. Hence
there is a SO$\left( d-2,2\right) $ covariance in $d$ dimensions. $d$ will
be determined in the quantum theory. It will be fixed to a critical
dimension $d_{crit}=27$ for $m_2\neq 0,$ or $d_{crit}=28$ for $m_2=0,$ to
eliminate quantum anomalies.

We will first solve the constraints classically and hence we will add $
J_n^{\pm }=0$ as a gauge choice to the massive case as well. This solution
will be used in a lightcone quantization. Using the SO$\left( d-2,2\right) $
symmetry one can boost $p_2^\mu $ to a timelike rest frame if the mass $
m_2\neq 0,$ or to a lightlike frame if the mass $m_2=0,$ 
\begin{equation}
\hat{p}_2^\mu =\left( m_2,0;\vec{0},0\right) \quad or\quad \hat{p}_2^\mu
=\left( \left| p_2\right| ,0;\vec{0},p_2\right) ,  \label{special}
\end{equation}
where the first two entries are timelike and the rest are spacelike. This
allows the solution of three constraints in the form 
\begin{eqnarray}
m_2 &\neq &0:\quad \hat{p}_{1\mu }=\left( 0,\hat{p}_{1I}\right) ,\quad \hat{
\alpha}_{\mu n}^{\pm }=\left( 0,\hat{\alpha}_{In}^{\pm }\right) ,
\label{sol} \\
m_2 &=&0:\quad \hat{p}_1^\mu =\left( 0,\hat{p}_{1I},0\right) ,\quad \hat{
\alpha}_{\mu n}^{\pm }=\left( 0,\hat{\alpha}_{In}^{\pm },0\right) , 
\nonumber
\end{eqnarray}
leaving one constraint to be solved 
\begin{equation}
L_n^{\pm }=\frac 12\sum_{m=-\infty }^\infty \hat{\alpha}_{n-m}^{\pm }\cdot 
\hat{\alpha}_m^{\pm }=0,  \label{sol2}
\end{equation}
where $\hat{\alpha}_{In}^{\pm }$ are the string variables in the rest frame
or lightcone frame of particle \#2, and the index $I$ labels a subspace of
signature $\left( -1,1,\cdots ,1\right) $ whose dimension is $\left(
d-1\right) $ for $m_2\neq 0$ and $\left( d-2\right) $ for $m_2=0.$ Therefore
there remains a Lorentz symmetry SO$\left( d-2,1\right) $ for $m_2\neq 0$
and SO$\left( d-3,1\right) $ for $m_2=0.$ Note that for $m_2=0,$ the
solutions $\hat{\alpha}_{\mu n}^{\pm },\hat{p}_1^\mu $ have no components
along the lightlike $\hat{p}_2^\mu .$ This is due to the gauge symmetry (\ref
{three}) as explained above.

\subsection{Covariant quantization}

To quantize covariantly and implement the constraints (\ref{covco}) on the
states one may use the BRST formalism. The BRST procedure is valid for first
class constraints. We will pretend as if it is applies to both massless and
massive cases with the constraints $\Phi ,$ $L_n^{\pm }\,,J_n^{\pm }$ in
order to illustrate the problem with the anomaly in (\ref{anom}), and will
find out that this set of constraints work only for $m_2=0.$ Then we will
discuss separately the massive case by dealing directly only with the first
class constraints (without the $J_n^{\pm }$).

Corresponding to the constraints $\Phi ,$ $L_n^{\pm }\,,J_n^{\pm }$ one
introduces the ghosts $\left( B,C\right) ,$ $\left( b_n^{\pm },c_n^{\pm
}\right) $ and $\left( \beta _n^{\pm },\gamma _n^{\pm }\right) .$ The
algebra of the constraints and their anomalies are 
\begin{eqnarray}
\left[ L_n^{\pm },L_m^{\pm }\right] &=&\left( n-m\right) L_{n+m}^{\pm }+
\frac d{12}\left( n^3-n\right) \delta _{n+m,0}  \nonumber \\
\left[ L_n^{\pm },J_m^{\pm }\,\right] &=&-m\,\,\,J_m^{\pm } \\
\left[ J_n^{\pm }\,,\,J_m^{\pm }\,\right] &=&n\,\Phi \delta
_{n+m}-n\,m_2^2\delta _{n+m}  \nonumber
\end{eqnarray}
The BRST operator is 
\begin{eqnarray}
Q &=&C\Phi +\sum_{\pm }\sum_n\left[ c_{-n}^{\pm }L_n^{\pm }\,+\gamma
_{-n}^{\pm }J_n^{\pm }\,\right]  \nonumber \\
&&-\frac 12\sum_{\pm }\sum_{n,m}\left( n-m\right) \,c_{-n}^{\pm }c_{-m}^{\pm
}b_{n+m}^{\pm } \\
&&-\frac 12\sum_{\pm }\sum_mm\gamma _{-m}^{\pm }\gamma _m^{\pm }B  \nonumber
\\
&&-\sum_{\pm }\sum_{n,m}\left( -m\right) \,c_{-n}^{\pm }\gamma _{-m}^{\pm
}\beta _{n+m}^{\pm }-\sum_{\pm }c_0^{\pm }\,a^{\pm }-CA  \nonumber
\end{eqnarray}
where $a^{\pm },A$ are anomaly constants to be determined by requiring $
Q^2=0.$ By anticommuting $Q$ with the antighosts $B,b_n,\beta _n$ one gets
the total gauge generators of the BRST quantized theory 
\begin{eqnarray}
\Phi ^{tot} &=&\Phi -A  \nonumber \\
L_n^{\left( \pm \right) tot} &=&L_n^{\pm }+\sum_m\left( n-m\right)
\,b_{n+m}^{\pm }c_{-m}^{\pm } \\
&&+\sum_m\left( -m\right) \,\beta _{n+m}^{\pm }\gamma _{-m}^{\pm }-a^{\pm
}\delta _{n0}  \nonumber \\
J_n^{\left( \pm \right) tot} &=&J_n^{\pm }+n\left( \gamma _n^{\pm
}B+\sum_m\beta _{n+m}^{\pm }c_{-m}^{\pm }\right)  \nonumber
\end{eqnarray}
Requiring $Q^2=0$ is equivalent to requiring the total generators to close
without anomalies. For a ghost system of dimension $h$ the Virasoro
generator is $L_n^{\left( h\right) }=\sum_m\left( n(h-1)-m\right)
\,b_{n+m}^{\left( h\right) }c_{-m}^{\left( 1-h\right) }$ and the anomaly in
its algebra is 
\begin{eqnarray}
anom\,(h) &=&\delta _{n+m}\frac 1{12}\left[ n^3c_h+2n\right] \\
c_h &=&-12h^2+12h-2.  \nonumber
\end{eqnarray}
Therefore the total anomaly in the algebra of $L_n^{\left( \pm \right) tot}$
is 
\begin{equation}
\frac{n^3}{12}(d+c_{h=2}+c_{h=1})+\frac n{12}(-d+24a^{\pm }+2+2)
\end{equation}
with the total central charge 
\begin{equation}
c_{tot}=d+c_{h=2}+c_{h=1}=d-28.
\end{equation}
The $J_n^{\left( \pm \right) tot}$ algebra is 
\begin{eqnarray}
\left[ J_n^{\left( \pm \right) tot}\,,\,J_m^{\left( \pm \right)
tot}\,\right] &=&n\,\Phi ^{tot}\,\delta _{n+m} \\
&&+n\,\left( A-m_2^2\right) \delta _{n+m},  \nonumber
\end{eqnarray}
and 
\begin{equation}
\left[ L_n^{\left( \pm \right) tot}\,,\,J_m^{\left( \pm \right)
tot}\,\right] =-mJ_{n+m}^{\left( \pm \right) tot}.
\end{equation}
So the total anomaly cancels if 
\begin{equation}
d=28,\quad a^{\pm }=1,\quad A=m_2^2.
\end{equation}
Under these conditions one finds 
\begin{equation}
\Phi ^{tot}=p_2^2+m_2^2-A=p_2^2=0.
\end{equation}
So, the quantum particle must be{\it \ massless} $p_2^2=0,\,$and the
critical dimension for the total system is 28. The global Lorentz symmetry
of this quantum system is SO$\left( 26,2\right) .$ Therefore, the analysis
applies correctly only to the $m_2=0$ case, as anticipated above.

For $m_2\neq 0$ the correct treatment of the second class constraint $
J_n^{\left( \pm \right) }=0$ needs more care. Since the present model is in
the class of gauged WZW models one may imitate the BRST procedure advocated
in \cite{schnitzer} to show that the final result is equivalent to 
the standard 
coset construction. In the coset language our case corresponds to $G/H$ 
with $G=R^d$ and $H=R,$ and our Virasoro generator 
in (\ref{covvir}) is indeed the
coset\ construction for the conformal field theory for $R^d/R.$ The central
charge of this Virasoro generator is $d-1,$ and for the quantum consistency
of the conformal field theory it must be set equal to 26. Hence 
\begin{equation}
m_2\neq 0:\quad d_{crit}=27,
\end{equation}
and the global symmetry of the system with a massive particle \#2 is SO$
\left( 25,2\right) .$

This result may be obtained more directly without appealing to the formalism
of \cite{schnitzer}. Namely, one can avoid the $J_n^{\left( \pm \right) }=0$
gauge fixing and work directly with the oscillators 
\begin{equation}
\tilde{\alpha}_{n\mu }^{\left( \pm \right) }=\alpha _{n\mu }^{\left( \pm
\right) }-\frac 1{p_2^2}p_{2\mu }\,\,p_2\cdot \alpha _n^{\left( \pm \right) }
\end{equation}
since these $\tilde{\alpha}_{n\mu }^{\left( \pm \right) }$ solve explicitly
the original constraints $p_2\cdot D_{\pm }x_1=0$. The Virasoro constraints
in (\ref{covvir}) are written directly in terms of these oscillators 
\begin{equation}
m_2=0:\quad L_n^{\pm }=\frac 12\sum_{m=-\infty }^\infty \tilde{\alpha}
_{n-m}^{\left( \pm \right) \mu }\cdot \tilde{\alpha}_m^{\left( \pm \right)
\nu }\,\,\eta _{\mu \nu },
\end{equation}
and their commutation rules are 
\begin{equation}
\left[ \tilde{\alpha}_n^{\left( \pm \right) \nu },\tilde{\alpha}_m^{\left(
\pm \right) \nu }\right] =n\,\delta _{n+m}\,\left( \eta ^{\mu \nu }-\frac{
p_2^\mu \,p_2^\nu }{p_2^2}\right) \,\,.
\end{equation}
The only constraints that need to be considered are the first class
constraints $L_n^{\pm }=0,\,\,\Phi =p_2^2+m_2^2=0,\,\,J_0=p_1\cdot p_2=0.$
In the algebra of these constraints all commutators are zero, except for 
\begin{eqnarray}
\left[ L_n^{\pm },L_m^{\pm }\right]  &=&\left( n-m\right) L_{n+m}^{\pm } \\
&&+\frac{d-1}{12}\left( n^3-n\right) \delta _{n+m,0}  \nonumber
\end{eqnarray}
where the $d-1$ anomaly comes from 
\begin{equation}
\left( \eta ^{\mu \nu }-\frac{p_2^\mu \,p_2^\nu }{p_2^2}\right) \left( \eta
_{\mu \nu }-\frac{p_{2\mu }\,p_{2\nu }}{p_2^2}\right) =d-1.
\end{equation}
The BRST operator is then 
\begin{eqnarray}
Q &=&\sum_{\pm }\sum_nc_{-n}^{\pm }L_n^{\pm }\,-\frac 12\sum_{\pm
}\sum_{n,m}\left( n-m\right) \,c_{-n}^{\pm }c_{-m}^{\pm }b_{n+m}^{\pm } 
\nonumber \\
&&+C\Phi +\gamma _0J_0-\sum_{\pm }c_0^{\pm }\,a^{\pm }
\end{eqnarray}
and the standard procedure gives the critical dimension $d-1=26.$

\subsection{Lightcone quantization}

The remaining constraints in the rest frame or lightcone frame of particle
\#2 are $(\partial _{\pm }\hat{x}_{1\pm }^I)^2=0$ or $L_n^{\pm }=0$ of eq.( 
\ref{sol2}). These are the familiar Virasoro constraints of string theory,
which can be solved explicitly by taking advantage of the conformal
invariance of the string system and choosing the lightcone gauge $\hat{x}
_1^{+}=\hat{p}_1^{+}\tau $ 
\begin{eqnarray}
\hat{\alpha}_n^{(\pm )-} &=&\frac 1{2\hat{p}_1^{+}}\sum_{m=-\infty }^\infty
\sum_I\hat{\alpha}_{n-m}^{(\pm )i}\,\hat{\alpha}_m^{(\pm )i}-\frac{\tilde{a}
}{\hat{p}_1^{+}}\delta _{n,0},  \label{lightcone} \\
\hat{\alpha}_n^{(\pm )+} &=&\hat{p}_1^{+}\,\delta _{n,0},\quad  \nonumber
\end{eqnarray}
The unconstrained degrees of freedom are the transverse string oscillators $
\vec{\alpha}_n^{(\pm )i}$ that describe the left/right moving string
excitations and the center of mass canonical degrees of freedom for the
string $(\hat{q}_1,\hat{p}_1)$ and particle $(\hat{q}_2,\hat{p}_2)$, all of
which are expressed in the rest frame or lightcone frame of particle \#2.

The Lorentz symmetry SO$\left( d-2,1\right) $ (or SO$\left( d-3,1\right) $ )
which was manifest in the special frame (\ref{special}-\ref{sol2}) is hidden
in the lightcone gauge (\ref{lightcone}) for the string. As is well known,
the quantum algebra for the {\it normal ordered} generators $\hat{M}
^{IJ}=\left( \hat{M}^{-+},\hat{M}^{i+},\hat{M}^{i-},\hat{M}^{ij}\right) $ 
\begin{eqnarray}
\hat{M}^{-+} &=&\hat{q}_1^{-}\hat{p}_1^{+},\quad \hat{M}^{i+}=\hat{q}_1^i
\hat{p}_1^{+},\quad  \nonumber \\
\hat{M}^{i-} &=&\hat{q}_1^i\hat{p}_1^{-}-\hat{q}_1^{-}\hat{p}_1^i-i\sum_{\pm
}\sum_{n\neq 0}\frac 1n:\hat{\alpha}_{-n}^{\left( \pm \right) i}\hat{\alpha}
_n^{\left( \pm \right) -}:,  \label{transv} \\
\hat{M}^{ij} &=&\hat{q}_1^i\hat{p}_1^j-\hat{q}_1^j\hat{p}_1^i-i\sum_{\pm
}\sum_{n\neq 0}\frac 1n:\hat{\alpha}_{-n}^{\left( \pm \right) i}\hat{\alpha}
_n^{\left( \pm \right) j}:,  \nonumber
\end{eqnarray}
of this symmetry closes correctly only if the number of transverse
dimensions labeled by $i$ is $24$ and $\tilde{a}=1$ (see e.g. \cite{GSW}).
The number of transverse dimensions is $\left( d-3\right) =24$ for $m_2\neq
0,$ and $\left( d-4\right) =24$ for $m_2=0$. Therefore the particle and
string system has a critical dimension 
\begin{eqnarray}
m_2 &\neq &0:\quad d_{crit}=27, \\
m_2 &=&0:\quad d_{crit}=28,  \nonumber
\end{eqnarray}
so the generators $\hat{M}^{IJ}$ represent correctly SO$\left( 25,1\right) $
at the quantum level.

The original action was invariant under the full classical rotation
invariance SO$\left( d-2,2\right) .$ For the critical dimension this
classical symmetry is SO$\left( 25,2\right) $ if $m_2\neq 0$ and SO$\left(
26,2\right) $ if $m_2=0.$ We now need to show that the quantum theory has
the higher symmetry by verifying that the Lorentz algebra closes. To
construct the remaining generators of SO$\left( 25,2\right) $ (or SO$\left(
26,2\right) $) we need to boost back to the general frame of particle \#2
and include the canonical degrees of freedom of particle \#2.

For the massive particle $m_2\neq 0$ in the general frame the mass shell
constraint is solved by the SO$\left( d-2,2\right) $ covariant vector 
\begin{equation}
p_2^\mu =\left( E_2^{\prime },p_2^I\right) ,\quad E_2^{\prime }=\sqrt{
p_2\cdot p_2+m_2^2}
\end{equation}
where $p_2^I$ is a SO$\left( 25,1\right) $ vector. The boost of any vector $
\hat{v}^\mu =\left( \hat{v}^{0^{\prime }},\hat{v}^I\right) $ defined in the
particle rest frame (denoted with the hats \symbol{94}) to the vector $v^\mu
=\left( v^{0^{\prime }},v^I\right) $ defined in the particle general frame
(no \symbol{94}) is given by 
\begin{eqnarray}
v^{0^{\prime }} &=&\frac 1{m_2}\left( E_2^{\prime }\hat{v}^{0^{\prime
}}+p_2\cdot \hat{v}\right) ,\quad  \nonumber \\
v^I &=&\hat{v}^I+\frac{p_2^I}{m_2}\left( \frac{p_2\cdot \hat{v}}{E_2^{\prime
}+m_2}+\hat{v}^{0^{\prime }}\right)
\end{eqnarray}
where the sum over $I$ in the dot products is SO$\left( 25,1\right) $
covariant in $26$ dimensions. Of course, the transformation is such that dot
products are SO$\left( 25,2\right) $ invariant in the full $27$ dimensions $
\hat{v}^2=v^2.$ The string and particle can now be described in the general
frame by boosting the rest frame solution in (\ref{special}-\ref{sol2}).
Taking into account $\hat{\alpha}_n^{\left( \pm \right) 0^{\prime }}=0$ one
obtains 
\begin{equation}
\alpha _n^{\left( \pm \right) 0^{\prime }}=\frac{p_2\cdot \hat{\alpha}
_n^{\pm }}{m_2},\quad \alpha _n^{\left( \pm \right) I}=\hat{\alpha}
_n^{\left( \pm \right) I}+\frac{p_2^I}{m_2}\frac{p_2\cdot \hat{\alpha}
_n^{\pm }}{E_2^{\prime }+m_2}\,\,.  \label{covalpha}
\end{equation}
These are expected to form covariant SO$\left( 25,2\right) $ vectors $\alpha
_n^{\left( \pm \right) \mu }=\left( \alpha _n^{\left( \pm \right) 0^{\prime
}},\alpha _n^{\left( \pm \right) I}\right) ,$ as will be verified below.
Furthermore, because of the SO$\left( 25,2\right) $ invariance of dot
products, the fully SO$\left( 25,2\right) $ covariant Virasoro constraints
in (\ref{covvir}) are equal to (\ref{sol2}) for any $p_2^\mu =(E_2^{\prime
},p_2^I)$. Therefore, the explicit solution of these constraints is given in
terms of only the 24 transverse oscillators in (\ref{lightcone}). Thus, the $
27$ components $\alpha _n^{\left( \pm \right) \mu }$ given in (\ref{covalpha}
) are also expressed in terms of the 24 oscillators in (\ref{lightcone})
which are the ones that solve all the constraints in the general frame of
the massive particle. So, for example the $0^{\prime }$ component is 
\begin{equation}
\alpha _n^{\left( \pm \right) 0^{\prime }}=\frac 1{m_2}\left( -p_2^{+}\cdot 
\hat{\alpha}_n^{\left( \pm \right) -}-p_2^{-}p_1^{+}\delta
_{n,0}+p_2^i\alpha _n^{\left( \pm \right) i}\right)
\end{equation}
where $\hat{\alpha}_n^{\left( \pm \right) -}$ is quadratic in the 24
transverse oscillators as given in (\ref{lightcone}).

We are now ready to construct the generators $M^{\mu \nu }=\left(
M^{0^{\prime }I},M^{IJ}\right) $ of SO$\left( 25,2\right) $ in the general
frame of the massive particle \#2. They are given by 
\begin{eqnarray}
M^{0^{\prime }I} &=&\frac 12\left( q_2^IE_2^{\prime }+E_2^{\prime
}q_2^I\right) +\frac{p_{2J}\hat{M}^{JI}}{E_2^{\prime }+m_2} \\
M^{IJ} &=&q_2^Ip_2^J-q_2^Jp_2^I+\hat{M}^{IJ}  \nonumber
\end{eqnarray}
where $\hat{M}^{IJ},$ which satisfy the SO$\left( 25,1\right) $ Lie algebra,
are given in terms of the $24$ transverse string oscillators in eq.(\ref
{transv}). It can be checked that these $M^{\mu \nu }$ satisfy the SO$\left(
25,2\right) $ Lie algebra without any anomalies, and furthermore, that they
rotate the $\alpha _n^{\left( \pm \right) \mu }$ of (\ref{covalpha}) as
vectors 
\begin{equation}
\left[ M^{\mu \nu },\alpha _n^{\left( \pm \right) \lambda }\right] =i\eta
^{\mu \lambda }\alpha _n^{\left( \pm \right) \nu }-i\eta ^{v\lambda }\alpha
_n^{\left( \pm \right) \mu }.
\end{equation}
This last property is trivial for the $M^{IJ},\alpha _n^{\left( \pm \right)
K}$ since it is the same as the usual 26 dimensional string in the lightcone
gauge. The new feature is the structure of $M^{0^{\prime }I}$. It can be
checked that this structure automatically closes into the higher algebra SO$
\left( 25,2\right) $ provided the $\hat{M}^{IJ}$ form the SO$\left(
25,1\right) $ Lie algebra.

The form of $M^{0^{\prime }I}$ follows from rather general properties of
cosets. The $\alpha _n^{\left( \pm \right) \mu }$ are given by boosting the $
\hat{\alpha}_n^{\left( \pm \right) \mu }$ with a $p_2^\mu $ dependent boost 
\begin{equation}
\alpha _n^{\left( \pm \right) \mu }=T_{\,\,\,\,\,\nu }^\mu \left( p_2\right) 
\hat{\alpha}_n^{\left( \pm \right) \nu }.
\end{equation}
where $T_{\,\,\,\,\,\nu }^\mu \left( p_2\right) $ is in the coset SO$\left(
25,2\right) /$SO$\left( 25,1\right) .$ When a general SO$\left( 25,2\right) $
transformation is applied, it can be rewritten as follows 
\begin{eqnarray}
\alpha _n^{\left( \pm \right) \mu } &\rightarrow &\Lambda _{\,\,\,\,\,\nu
}^\mu \,\alpha _n^{\left( \pm \right) \nu }=\left( \Lambda T(p_2)\right)
_{\,\,\,\,\,\nu }^\mu \hat{\alpha}_n^{\left( \pm \right) \nu } \\
&=&\left( T(p_2^{\prime })H\right) _{\,\,\,\,\,\nu }^\mu \hat{\alpha}
_n^{\left( \pm \right) \nu }  \nonumber
\end{eqnarray}
where $p_2^{\prime \mu }=\Lambda _{\,\,\,\,\,\nu }^\mu p_2^\nu ,$ and $
H\left( p_2,\Lambda \right) $ is an element in the subgroup SO$\left(
25,1\right) \,\,$but its parameters depend on a function of both $p_2^\mu $
and $\Lambda _{\,\,\,\,\nu }^\mu .$ When $\Lambda $ is an element of SO$
\left( 25,1\right) $ one has $H=\Lambda $, therefore the subgroup is
implemented on the 26 $\hat{\alpha}_n^{\left( \pm \right) I}$ and $p_2^I$ by
the {\it total} particle and string generators $M^{IJ}.$ Of course, the $
\hat{\alpha}_n^{\left( \pm \right) 0^{\prime }}$ is invariant under this
combined transformation since it is a dot product. The remaining SO$\left(
25,2\right) /$SO$\left( 25,1\right) $ coset transformations have generators
that are precisely the $M^{0^{\prime }I}$ given above, and they
automatically take into account the complicated nature of $H\left(
p_2,\Lambda \right) .$

The outcome of the lightcone quantization for $m_2\neq 0$ is a critical
dimension $d=27$ with signature $\left( 25,2\right) ,$ in agreement with the
covariant quantization.

The lightcone quantization for the $m_2=0$ case can be done in a similar
way. One needs to boost back from the lightcone frame of particle \#2 to the
general frame. It is a straightforward exercise and there are no oscillator
ordering problems, just as in the massive case. Therefore the critical
dimension is $d=28$ with signature $\left( 26,2\right) $, in agreement with
the BRST quantization.

\section{Multi particles, strings, p-branes}

The type of action discussed in this paper can be generalized to other
systems. For example for three particles 
\begin{equation}
S=S_1+S_2+S_3+\lambda _1\cdot \lambda _2+\lambda _2\cdot \lambda _3+\lambda
_3\cdot \lambda _1
\end{equation}
where $S_1\left( x_1^\mu ,A_{12},A_{13},e_1,\lambda _2^\mu ,\lambda _3^\mu
\right) $ is the action for particle \#1 in the background of particles
\#2,3, constructed in terms of gauge covariant derivatives 
\begin{equation}
D_\tau x_1^\mu =\partial _\tau x_1^\mu -A_{12}\lambda _2^\mu -A_{13}\lambda
_3^\mu
\end{equation}
in the spirit of gauged WZW models. So, the coset is $R^d/R^2.$ Similarly
for the actions $S_{2,3}$ which are obtained by a cyclic permutation of the
indices 1,2,3. Because of the gauge invariances one finds constraints, and
going through a similar analysis as the two particle case one determines $
\lambda _i^\mu \sim p_i^\mu $ and the constraints 
\begin{equation}
p_i\cdot p_j+m^2_i\delta _{ij}=0.
\end{equation}
The solution of this system of constraints requires 3 timelike coordinates.
In the bosonic case there seems to be no limit on the number of particles
and corresponding new timelike dimensions, but with supersymmetry there are
hints for both sufficient and necessary reasons to have a minimum as well as
a maximum of three timelike dimensions in a setting that is SO$\left(
11,3\right) $ covariant \cite{d14} . The structures of \cite{d14} were found
to be necessary and sufficient to unify type-A and type-B supersymmetries.
This unification is possible with a minimum of three timelike dimensions and
extending the general structure beyond 14 dimensions is not required by any
known phenomena. Furthermore, there seems to be an obstruction to Yang-Mills
supersymmetric systems beyond 14 dimensions \cite{sezgin3}, thus providing a
hint for a maximum of three time like dimensions.

The same approach can be applied to a string and two particles, with results
that can be guessed from the previously discussed cases, of a string \& one
particle, and three particles. The general result is that when a massive
particle is added one needs to add $\left( 0,1\right) $ dimensions, i.e. one
timelike dimension but no spacelike dimensions, and when a massless particle
is added one must add $\left( 1,1\right) ,$ i.e. one time plus one space
dimensions. Similarly one may substitute a membrane for a string, and so on
for other p-branes, in the case of classical p-branes. Of course, one does
not know how to solve the quantum theory that includes membranes or
p-branes, and therefore there is no reliable statement on the number of
dimensions for which the quantum theory is consistent. However, there are
partial hints that $d_{crit}=11$ for supermembranes \cite{bsp}, therefore
for the combined supermembrane \& superparticle system, one may extrapolate
these hints to $d=12$ with signature $\left( 10,2\right) $ if the
superparticle is massive or $d=13$ with signature $\left( 11,2\right) $ if
the superparticle is massless. The supersymmetry of such a system is not
standard, as discussed in \cite{spartstring} and the next section.

Next we consider two strings, with an action $S=S_1+S_2+\lambda _1\cdot
\lambda _2$, where both $S_{1,2}$ are string actions of the form (\ref
{partstr}). The equations of motion and constraints for each string can be
solved, both classically and quantum mechanically, following the same steps
as sections 3,4. But now we find new features in the quantum consistency of
the combined system. Recall that in consistent sectors the masses of each
string are given by $-p_i^2=N_i-1,$ where $N_i$ are the oscillator
excitation levels. Perhaps the simplest way to arrive at the critical
dimension is to note that a quantum consistent string must have 26
dimensions after putting the other string in one of its massive or massless
states. If the second string is in a massless state, its effect on the first
string is the same as a massless particle. Then the total number of
dimensions for string \#1 has to be 28 with signature $\left( 26,2\right) $,
as we have shown in sections 3,4. On the other hand, if the second string is
in a massive state, its effect on the first string is the same as a massive
particle, and the number of dimensions must be 27, with signature $\left(
25,2\right) .$ There is also a third case when the mass of the second string
is tachyonic. Then the number of dimensions is also 27, but with signature $
\left( 26,1\right) .$ The roles of the two strings may be reversed and
similar statements would be made for the critical dimension of string \#2.
These statements cannot be all simultaneously right in the same theory,
since the classical theory is defined with a fixed number of dimensions for
both strings. Hence, as the consistent quantum sectors, we must select only
the mass sectors that are simultaneously consistent for both strings for a
fixed number of dimensions. The sectors are defined by whether the masses $
(-p_1^2,\,-p_2^2)\,\,$are simultaneously zero, positive or negative, and
evidently the only consistent sectors are 
\begin{eqnarray}
\left( 26,2\right) &:&\quad (-p_1^2,\,-p_2^2)=\left( 0,0\right)  \nonumber \\
\left( 25,2\right) &:&\quad (-p_1^2,-\,p_2^2)=\left( +,+\right) \\
\left( 26,1\right) &:&\quad (-p_1^2,-\,p_2^2)=\left( -,-\right)  \nonumber
\end{eqnarray}
The $\left( 0,0\right) $ sector which is possible in 28 dimensions has only
one state, similarly the $\left( -,-\right) $ sector in 27 dimensions has
only one state, while the $\left( +,+\right) $ sector in 27 dimensions has
an infinite number of massive states from each string. Here we have assumed
that the conformal field theory for each string has no spectator sectors,
that is that all degrees of freedom of both strings are coupled via the
coupling $\lambda _1\cdot \lambda _2.$ Of course, this assumption can be
modified by changing the model.

Consider a model that has one dimension for string \#1 which remains
uncoupled, while all other $\left( d_1-1\right) $ dimensions of string \#1
are coupled to the $d_2$ dimensions of string \#2 via $\lambda _1\cdot
\lambda _2\sim p_1\cdot p_2$, so that $d_2=\left( d_1-1\right) .$ So, string
\#1 has one extra dimension. We need to consider again the values of $
(-p_1^2,-\,p_2^2).$ The mass shell conditions in consistent sectors are now $
-p_1^2+\tilde{p}^2=N_1-1$ and $-p_2^2=N_2-1,$ where $\tilde{p}^2$ is the
zero mode of the extra dimension of string \#1. If this dimension is compact
there would be contributions from the winding sectors as well. In this model
the sector $(-p_1^2,-\,p_2^2)=\left( +,0\right) $ is consistent for $d_1=28$
with signature $\left( 26,2\right) $ and $d_2=27$ with signature $\left(
25,2\right) .$ There are an infinite number states from string \#1 and only
one state from string \#2. Evidently, one can construct various consistent
models provided the sectors are selected as above.

However, it is not clear that such sectors are self consistent by themselves
under interactions. It is not yet clear what interactions should be
considered. If all interactions defined through vertex operator products of
both strings are included, then the sectors identified above do not seem to
remain isolated from others. Perhaps one can make sense of interactions that
mix sectors of different dimensions and signatures. More study is required
to understand such issues. These questions did not arise for the string \&
particle or string \& two particle systems.

\section{Supersymmetry}

In a separate publication the supersymmetric version of the superstring and
a massive or massless superparticle will be discussed in detail \cite
{spartstring}. This involves a construction of an action for the massive
superparticle and a more general superstring action that is invariant under
a generalized supersymmetry. Here we wish to mention some generalities and
ideas for future applications and improvements. By extrapolating from the
results of the present paper to the supersymmetric case, one expects
critical dimensions $d=12$ for $m_2=0$ and $d=11$ for $m_2\neq 0$ with a
Lorentz symmetry SO$\left( 10,2\right) $ and SO$\left( 9,2\right) $
respectively. Furthermore, as expected quite generally from \cite{stheory},
and from discussions in \cite{superpv,twotimes,sezgin3}, the generalized
superalgebra has to be 
\begin{equation}
\left\{ Q_\alpha ,Q_\beta \right\} =\gamma _{\alpha \beta }^{\mu \nu
}\,\,p_{1\mu }\,p_{2\nu }{,}  \label{newsusy}
\end{equation}
where $Q_\alpha $ is the Majorana-Weyl spinor of SO$\left( 10,2\right) $ or
SO$\left( 9,2\right) $ with 32 real components. In the lightcone frame of
the massless particle \#2 (as well as in the rest frame of the massive
particle \#2), the remaining Lorentz symmetry is SO$\left( 9,1\right) $ and
the supersymmetry reduces to the standard form of the ten dimensional type
IIA. Furthermore, the superstring reduces to the usual 10 dimensional type
IIA string. By contrast, in the general frame, or in the action, there is
full covariance under SO$\left( 10,2\right) $ or SO$\left( 9,2\right) .$

We believe that the massless system $m_2=0,\,$underlies a supergravity
theory in 12 dimensions, with bilocal fields $\Phi \left( x_1,x_2\right) $
describing the string-particle system, along the lines first suggested in 
\cite{stheory}. This supergravity theory has been partially realized in the
special lightcone frame in \cite{nishino}, but in a single Kaluza-Klein mode
of the bilocal fields $\Phi \left( x_1,x_2\right) $ (that is particle \#2
has been frozen to be at a fixed momentum $p_2^\mu )$. Similar
considerations for two particles (rather than string and particle) underlie
the Yang-Mills theories in 12 dimensions, which have also been only
partially realized in a similar Kaluza-Klein mode \cite{sezgin}.

As has been argued in \cite{d14} the unification of type-A and type-B
supersymmetries point to a unifying supersymmetric structure in 14
dimensions with signature $\left( 11,3\right) $. Some sectors of such a
structure can be constructed by considering three superparticles, or a
superstring with two superparticles, etc.. It is expected from \cite{d14},
and it has been confirmed in \cite{sezgin3}, that the three superparticle
system underlies a super Yang-Mills theory in 14 dimensions. This theory has
been partially constructed \cite{sezgin2} in a Kaluza-Klein sector in the
same sense as the two particle case (i.e. the sector in which the momenta of
two particles out of three are frozen). A superstring and two superparticles
probably underlie a supergravity in 14 dimensions that would generalize \cite
{nishino} to 14 dimensions. A more general approach that includes all
Kaluza-Klein modes has been illustrated in \cite{superpv} for free fields.
This approach needs to be further developed to include interactions by
figuring out the calculus of representations of the new supersymmetry (\ref
{newsusy}). It could then be applied to the construction of the full 12 or
14 dimensional supergravity and super Yang-Mills theories in all
Kaluza-Klein sectors.

\section{Acknowledgments}

I.B. was on sabbatical leave from the Department of Physics and Astronomy,
University of Southern California, Los Angeles, CA 90089-0484, USA. His
research was partially supported by a DOE grant No. DE-FG03-84ER40168, and
by USC.

C.K. was on leave from Ecole Normale Sup\'erieure, 24 rue Lhomond, F-75231, Paris, Cedex 05, France. His research was partially supported by EEC grant No. ERB-FMRX-CT 96/0045.

\vfill\eject

\end{document}